\RequirePackage{fixltx2e}
\RequirePackage{fix-cm}

\documentclass[%
showkeys,
twocolumn, %
nofootinbib, %
superscriptaddress, %
nofootinbib, %
]{revtex4-1}

\usepackage{cmap}

\usepackage{ucs}
\usepackage[utf8x]{inputenc}
\usepackage[T1,T2A]{fontenc}
\usepackage[german,russian,english]{babel}

\usepackage[sort&compress]{natbib}
\usepackage{amsmath}
\usepackage{amssymb}

\usepackage{mathtools}
\mathtoolsset{
showonlyrefs,
mathic = true
}

\allowdisplaybreaks

\usepackage{hyperref}
\hypersetup{backref,
 colorlinks=false}
\hypersetup{pdfborder=0 0 0}

\usepackage{breakurl}

\usepackage{microtype}
\UseMicrotypeSet[protrusion]{alltext}

\usepackage{fixltx2e}

\usepackage{nicefrac}

\makeatletter
\def\ps@pprintTitle{%
     \let\@oddhead\@empty
     \let\@evenhead\@empty
     \let\@oddfoot\@empty
     \let\@evenfoot\@oddfoot}
\makeatother

\usepackage{comment}

\usepackage{physics}

\renewcommand{\i}{\ensuremath{\mathrm{i}}}

\renewcommand{\Re}{\mathop{\mathrm{Re}}}

\newcommand{\crd}[1]{{\underline{\vphantom{j}{#1}}}}

\usepackage{tensor}

\begin{document}
\graphicspath{{image/geom-maxwell-cas/}}

\title{Using Two Types of Computer Algebra Systems to Solve Maxwell Optics Problems}

\author{D. S. Kulyabov}
\email{yamadharma@gmail.com}
\affiliation{Department of Applied Probability and Informatics\\
Peoples' Friendship University of Russia\\
Miklukho-Maklaya str. 6, Moscow, 117198, Russia}
\affiliation{Laboratory of Information Technologies\\
Joint Institute for Nuclear Research\\
Joliot-Curie 6, Dubna, Moscow region, 141980, Russia}

\thanks{Published in:
\textit{D.~S. Kulyabov,}
  \href{http://link.springer.com/10.1134/S0361768816020043}{{Using two types of
  computer algebra systems to solve maxwell optics problems}}, Programming and
  Computer Software 42~(2) (2016) 77--83.
\href {http://dx.doi.org/10.1134/S0361768816020043}{\path{doi:10.1134/S0361768816020043}}.
}

\thanks{Sources:
  \url{https://bitbucket.org/yamadharma/articles-2014-geom-maxwell-isotropic}}

\begin{abstract}







To synthesize Maxwell optics systems, the mathematical apparatus of
tensor and vector analysis is generally employed. This mathematical
apparatus implies executing a great number of simple stereotyped
operations, which are adequately supported by computer algebra
systems. In this paper, we distinguish between two stages of working
with a mathematical model: model development and model usage. Each of
these stages implies its own computer algebra system. As a model
problem, we consider the problem of geometrization of Maxwell's
equations. Two computer algebra systems---Cadabra and FORM---are
selected for use at different stages of investigation.

\end{abstract}


\maketitle

\section{Introduction}

This paper considers the application of computer algebra systems to
designing a Maxwell optics system that is described by Maxwell's
equations in arbitrary locally-orthogonal curvilinear coordinates. To
describe this problem mathematically using the apparatus of tensor and
vector analysis, computer algebra systems that support tensor calculus
are required. An additional requirement is the freewareness of such
computer algebra systems, since, at the current stage of
investigation, this problem is of scientific interest and involves no
commercial benefit. Among freeware computer algebra systems, which
implement the apparatus of tensor and vector analysis, two different
systems are selected for use at two different stages of
investigation~\cite{hadamard:psychology::en}. 

The first stage consists in writing a prototype program; for this
purpose, the Cadabra system can be successfully used. The second stage
is of particular importance for the domain experts who run the
constructed program. At this stage, a great deal of trial computations
with different parameters, which are arbitrary or systematically
varied, are carried out to find the optimum solution among feasible
ones or even a new solution of the problem under consideration. The
FORM system seems to be the best choice for this stage. 

Our attempts to find a ``silver
bullet''~\cite{brooks:1986:silver_bullet} have not been successful,
and searching for such a universal system has led us to the conclusion
that each problem requires its own specific remedy.

This paper is organized as follows. Section~\ref{sec:notation} introduces basic
notations and conventions. Types of computer algebra systems are
considered in
Section~\ref{sec:cas-type}. Section~\ref{sec:maxwell-geom} describes
the formalism of
geometrization of Maxwell's equations. Section~\ref{sec:maxwell:calc}
illustrates the use
of this formalism by an example of designing and calculating
three-dimensional waveguide objects of Maxwell optics. 

\section{Notations and conventions}
\label{sec:notation}

\begin{enumerate}
\item  In this paper, the abstract index notation is
used~\cite{penrose-rindler:spinors::en}, in which a
tensor, as a complete object, is denoted simply by an index (for
example, $x^{i}$), while its components are denoted by an underlined index
(for example, $x^{\crd{i}}$). 

\item  We adhere to the following conventions. Greek indices ($\alpha$
  and $\beta$) refer
to a four-dimensional space and, in a component form, have the
following notation: $\crd{\alpha} = \overline{0,3}$. Latin indices
from the middle of the alphabet 
($i$, $j$, and $k$) refer to a three-dimensional space and, in a component form,
have the following notation: $\crd{i} = \overline{1,3}$. 

\item  In the index, a comma denotes a partial derivative with respect to
the corresponding coordinate ($f_{,i} := \partial_{i} f$), while a
semicolon denotes a covariant
derivative ($f_{;i} := \nabla_{i} f$). 

\item  To write electrodynamics equations, the symmetric CGS system is
used. 
\end{enumerate}

\section{Types of computer algebra systems}
\label{sec:cas-type}

Computer algebra systems can be classified using various
criteria. Here, we confine ourselves to the criterion of
interactivity. Originally, batch processing prevailed in computing
systems. However, when the power of computers had made it possible to
reduce the response time to an acceptable level, interactivity became
a leading paradigm. Each paradigm has its own area of application. In
software development, interactivity was associated with prototyping
tools, while the classical approach (programming, compilation,
debugging, etc.) was used to create software products. In symbolic
computation systems, interactivity became a ruling principle. Indeed,
computer algebra systems were designed to increase the labor
productivity of scientists and served as a kind of a smart
notebook. Little by little, the problems involving noninteractive
computations were pushed to the sidelines together with the
corresponding computer algebra systems. 

The architecture of many modern computer algebra systems became an
obstacle in the way of increasing the computational
efficiency. Therefore, the approach that implies using different
computer algebra systems to solve different problems seems promising. 

In this paper, we consider two extreme poles. On the one side is the
Cadabra system, which is used to manipulate abstract objects. In this
case, the human--computer interaction is mandatory. On the other side
is a completely noninteractive (even batch) system, which resembles
(in terms of development cycle) classical programming languages rather
than common computer algebra systems.

\subsection{Cadabra}

The Cadabra is a special-purpose computer algebra system (more
information is available on \url{http://cadabra.phi-sci.com}). It is mainly
oriented to solving field theory problems. Since complex tensor
computations are an integral part of the field theory, it is no wonder
that this system support tensor computations to a high standard~\cite{kulyabov:2013:springer:cadabra,
  kulyabov:2009:nucleilett:cadabra,
  peeters:2006:field-theory_approach,
  peeters:2007:introducing_cadabra,
  peeters:2007:symbolic_field_theory, Brewin2010}.
The Cadabra system extensively uses the notation of the \TeX{}
typesetting system.\footnote{The deep integration between the Cadabra
  and \TeX{} resulted in the fact
that, upon installing \TeX{}Live-2015 (\url{http://www.tug.org/texlive/}), the
Cadabra terminated with an error: \texttt{Undefined control sequence:
\textbackslash{}int\_eval:w.} It was found that the error was caused by the conflict
between the \texttt{breqn} and \texttt{expl3} packages of \TeX{}Live-2015. The problem was
resolved by updating the \texttt{breqn} package from the CTAN repository
(\url{http://ctan.org}).}

Presently, the Cadabra system implements only operations with abstract
indices. Component computations are not supported. Yet, to implement
component computations, it is required to supplement the Cadabra
system with general-purpose tools of computer algebra.

\subsection{Form}

The FORM computer algebra system stands out quite markedly against the
background of the other like systems: it is oriented to batch
processing rather than to user interaction (more information is
available on \url{http://www.nikhef.nl/form})~\cite{tung:2005:FORM, vermaseren:2013:FORM_refman,
  heck:2000:formbook}. Hence, this system
does not suffer from certain inherent drawbacks of common computer
algebra systems such as high resource usage, restrictions on the
amount of computations, and slow speed. The FORM system supports
various technologies of parallel and distributed
computing~\cite{fliegner:1999:FORM} like,
for example, multithreading and several implementations of the message
passing interface (MPI). The system has an interface for interacting
with external programs~\cite{tentyukov:2007:extension_form}. Thus, the FORM is often used as a backend
to other (mostly, interactive) computer algebra systems. It finds an
especially wide
application in quantum field computations~\cite{boos:2010:automatic_calculation::en,
  bunichev:2003:comphep+form, hahn:1999:FeynArts, hahn:2000:report,
  hahn:2007:FeynEdit}.
The FORM system had been developed in 1984, but it was made open-source only in 2010. 

Basic features of the FORM are as follows: 
\begin{itemize}
\item arbitrary long mathematical expressions (limited only by the disk space); 
\item multithreaded execution and parallelization (MPI); 
\item fast trace calculation (of $\gamma$ matrices); 
\item output into various formats (text, Fortran, etc.); 
\item interface for communicating with external programs. 
\end{itemize}

\subsection{Comparison between Cadabra and FORM}

Below, the features of both the computer algebra systems, which are of
interest in this work, are outlined. 

\paragraph{Cadabra}
\begin{itemize}
\item The main feature is natural operations with tensors. Covariant and
contravariant indices are supported by default. 

\item The system is effective for writing new formulas and relations in an
interactive mode. 
\end{itemize}

\paragraph{FORM}
\begin{itemize}
\item The support of covariant and contravariant indices is implemented in
an artificial way. 

\item The system is effective for final computations with already known
formulas. 
\end{itemize}

Thus, the Cadabra system is more appropriate for writing new formulas
(see Section~\ref{sec:maxwell-geom}), while the FORM system is more suitable for direct
computations with already derived formulas (see Section~\ref{sec:maxwell:calc}).

\section{Geometrization of Maxwell's equations}
\label{sec:maxwell-geom}

The two computer algebra systems described above are used for computer
modeling and synthesis of Maxwell optics elements in terms of tensor
calculus and curvilinear coordinates. This problem is naturally solved
in two stages. First, the Cadabra system is used, since it is designed
to manipulate tensor objects. All manipulations are performed with
abstract indices only. Using the computer algebra system at this stage
makes it possible to get rid of paper and a pen when carrying out the
theoretical work. At the second stage, the FORM system is used to
create a software complex for batch solution of stereotyped
problems. This system is oriented to vector and tensor analysis and
uses the noninteractive (batch) approach and parallelization of
computations and external memory. 

\subsection{Idea of geometrization}

The employed methodology of modeling and synthesis of Maxwell optics
elements uses an effective geometric
paradigm~\cite{wheeler:neutrinos::en}, in which certain
field parameters are translated into geometric parameters. In this
case, macroscopic parameters of a medium are geometrized. Thus, direct
and inverse problems can be solved: trajectories of electromagnetic
wave propagation are found from known macroscopic parameters or these
parameters are restored based on certain trajectories. 

I.~E.~Tamm was first to try using methods of differential geometry in
electrodynamics~\cite{tamm:1924:jrpc::en, tamm:1925:jrpc::en,
  tamm:1925:mathann}. In 1960, J.~Plebanski proposed a method for
geometrizing constitutive equations of an electromagnetic field
~\cite{Plebanski1960, Felice1971, Leonhardt2008,
  leonhardt:2009:light}; this method became classical. 

The basic idea of a naive geometrisation of Maxwell's equations
consists in the following. 
\begin{enumerate}
\item Write Maxwell's equations in a medium in the Minkowski space.
\item Write Maxwell's equations in a vacuum in the effective Riemannian space.
\item Equate the corresponding terms of the equations. 
\end{enumerate}
Thus, we obtain the expression of dielectric permittivity and magnetic
permeability via the metric of the corresponding effective space. 

In more detail, this technique is described in~\cite{kulyabov:2012:vestnik:2012-1,
  kulyabov:2013:springer:cadabra, kulyabov:2013:conf:maxwell,
  kulyabov:2011:vestnik:curve-maxwell::en}.

\subsection{General relations}

Recall some basic facts about Maxwell's equations. 

Write Maxwell's equation via the electromagnetic field tensors
$F_{\alpha\beta}$ and
$H_{{\alpha}{\beta}}$~\cite{minkowski:1908,stratton:1948::en}:
\begin{gather}
  \nabla_{{\alpha}} F_{{\beta}{\gamma}}+ \nabla_{{\beta}}
  F_{{\gamma}{\alpha}}+\nabla_{{\gamma}} F_{{\alpha}{\beta}} =
  F_{[\alpha \beta ; \gamma]} = 0,
  \label{eq:m:tensor:2}
  \\
  \nabla_{{\alpha}} H^{{\alpha}{\beta}}=\frac{4 \pi}{c}J^{{\beta}}.
  \label{eq:m:tensor}
\end{gather}
Here, the tensors $F_{\alpha\beta}$ and $H^{\alpha\beta}$ have the following components: 
\begin{gather}
  F_{\crd{\alpha}\crd{\beta}}=
  \begin{pmatrix}
    0 & {E}_1 & {E}_2 & {E}_3 \\
    -{E}_1 & 0 & -{B}^3 & {B}^2 \\
    -{E}_2 & {B}^3 & 0 & -{B}^1 \\
    -{E}_3 & -{B}^2 & {B}^1 & 0
  \end{pmatrix},
  \label{eq:f_ab}
  \\
  H^{\crd{\alpha}\crd{\beta}}=
  \begin{pmatrix}
    0 & -{D}^1 & -{D}^2 & -{D}^3 \\
    {D}^1 & 0 & -{H}_3 & {H}_2 \\
    {D}^2 & {H}_3 & 0 & -{H}_1 \\
    {D}^3 & -{H}_2 & {H}_1 & 0
  \end{pmatrix}.
  \label{eq:g^ab}
\end{gather}

To take into account the medium, introduce some macroscopic equations:
\begin{equation}
  \label{eq:constraint}
  D^{i} = \varepsilon^{i j} E_{j}, \qquad B^{i} = \mu^{i j} H_{j},
\end{equation}
\begin{equation}
  \label{eq:constraint:4}
  H^{\alpha \beta} = \lambda^{\alpha \beta}_{\gamma \delta} F^{\gamma
    \delta}, 
  \quad 
  \lambda^{\alpha \beta}_{\gamma \delta} = \lambda^{[\alpha \beta]}_{[\gamma \delta]}.
\end{equation}

Obviously, in the vacuum (when there is no medium),
relations~\eqref{eq:constraint} and~\eqref{eq:constraint:4}
take the following form: 
\begin{equation}
  \label{eq:perv:vac}
  \varepsilon^{i j} := \delta^{i j}, \qquad \mu^{i j} := \delta^{i j}.
\end{equation}
\begin{equation}
  \label{eq:constraint:vac}
  D^{i} = E^{i}, \qquad B^{i} = H^{i}, 
  \qquad H^{\alpha \beta} = F^{\alpha \beta}.
\end{equation}
  
\subsection{Geometrization of Maxwell's equations in Cartesian coordinates }

Note that, in differential Bianchi identity~\eqref{eq:m:tensor:2},
covariant derivatives can be replaced by partial derivatives: 
\begin{equation}
  \label{eq:m:tensor:2:partial}
  \begin{gathered}
    \nabla_{{\alpha}} F_{{\beta}{\gamma}}+ \nabla_{{\beta}}
    F_{{\gamma}{\alpha}}+\nabla_{{\gamma}} F_{{\alpha}{\beta}} =
    F_{[\alpha \beta ; \gamma]} = 0,
    \\
    \Downarrow
    \\
    \partial_{{\alpha}} F_{{\beta}{\gamma}}+ \partial_{{\beta}}
    F_{{\gamma}{\alpha}}+\partial_{{\gamma}} F_{{\alpha}{\beta}} =
    F_{[\alpha \beta , \gamma]} = 0.
  \end{gathered}
\end{equation}
Indeed,
\begin{multline}
  \label{eq:m:tensor:2:detail}
  \nabla_{{\alpha}} F_{{\beta}{\gamma}}+ \nabla_{{\beta}}
  F_{{\gamma}{\alpha}}+\nabla_{{\gamma}} F_{{\alpha}{\beta}} = {} \\
  {} =
  \partial_{{\alpha}} F_{{\beta}{\gamma}} -
  \Gamma_{\alpha\beta}^{\delta} F_{\delta \gamma} - \Gamma_{\alpha
    \gamma}^{\delta} F_{\beta \delta} + {} \\ {} +
  \partial_{{\beta}} F_{{\gamma}{\alpha}} - \Gamma_{\beta
    \gamma}^{\delta} F_{\delta \alpha} - \Gamma_{\beta
    \alpha}^{\delta} F_{\gamma \delta} + {} \\ {} +
  \partial_{{\gamma}} F_{{\alpha}{\beta}} - \Gamma_{\gamma
    \alpha}^{\delta} F_{\delta \beta} - \Gamma_{\gamma
    \beta}^{\delta} F_{\alpha \delta}.
\end{multline}

Let us demonstrate the evaluation of
expression~\eqref{eq:m:tensor:2:detail} in the Cadabra
system.\footnote{A point at the end of an instruction suppresses the
  output of the result, while a semicolon allows printing the result.}

Set a list of indices:
\begin{verbatim}
{\alpha,\beta,\gamma,\delta}::Indices(vector).
\end{verbatim}

Introduce partial and covariant derivatives:\footnote{The symbol \# is
  a pattern for any expression.} 
\begin{verbatim}
\partial_{#}::PartialDerivative.
\end{verbatim}
\begin{verbatim}
\nabla_{#}::Derivative.
\end{verbatim}

When deriving expression~\eqref{eq:m:tensor:2:detail}, it is important
to take into account the symmetry of Christoffel symbols, which are
defined using a Young diagram~\cite{fulton:young_tableaux::en}:
\begin{verbatim}
\Gamma^{\alpha}_{\beta \gamma}::
  TableauSymmetry(shape={2}, indices={1,2}).
\end{verbatim}

For the tensor $F_{\alpha \beta}$, there is no need in the Young diagram, it is
sufficient to state that this tensor is antisymmetric:
\begin{verbatim}
F_{\alpha \beta}::AntiSymmetric.
\end{verbatim}

Write the expression for the covariant derivative in the form of a
substitution:\footnote{The symbol $:=$ is used to set a label.}
\begin{verbatim}
nabla:=\nabla_{\gamma} A?_{\alpha \beta} -> 
  \partial_{\gamma}{A?_{\alpha \beta}} - 
  A?_{\alpha \delta}
  \Gamma^{\delta}_{\beta \gamma} - 
  A?_{\delta \beta}
  \Gamma^{\delta}_{\alpha \gamma};
\end{verbatim}
\begin{equation*}
  nabla:= {\nabla}_{\gamma}  {A?}_{\alpha \beta} \rightarrow
  ({\partial}_{\gamma}{{A?}_{\alpha \beta}}  - {A?}_{\alpha \delta}
  {\Gamma}^{\delta}_{\beta \gamma} - {A?}_{\delta \beta}
  {\Gamma}^{\delta}_{\alpha \gamma});
\end{equation*}
We use the postfix ``?'' to convert the preceding letter into a pattern
(otherwise, the substitution can be used only with a fixed
variable). Note that it is not necessary to apply the modifier ``?'' to
indices.

Now, rewrite equation~\eqref{eq:m:tensor:2} as follows:
\begin{verbatim}
maxwell1:= \nabla_{\alpha}F_{\beta \gamma} + 
  \nabla_{\beta} F_{\gamma \alpha} + 
 \nabla_{\gamma} F_{\alpha \beta};
\end{verbatim}
\begin{equation*}
maxwell1:= {\nabla}_{\alpha}  {F}_{\beta \gamma} +
{\nabla}_{\beta}  {F}_{\gamma \alpha} + {\nabla}_{\gamma}
{F}_{\alpha \beta};
\end{equation*}

Substitute \verb|nabla| into expression \verb|maxwell1|:
\begin{verbatim}
@substitute!(maxwell1)(@(nabla));
\end{verbatim}
\begin{multline*}
maxwell1:= {\partial}_{\alpha}{{F}_{\beta \gamma}}  - {F}_{\beta
  \delta} {\Gamma}^{\delta}_{\gamma \alpha} - {F}_{\delta \gamma}
{\Gamma}^{\delta}_{\beta \alpha} 
+ {} \\ {} + 
{\partial}_{\beta}{{F}_{\gamma
    \alpha}}  - {F}_{\gamma \delta} {\Gamma}^{\delta}_{\alpha \beta} -
{F}_{\delta \alpha} {\Gamma}^{\delta}_{\gamma \beta} 
+ {} \\ + {}
{\partial}_{\gamma}{{F}_{\alpha \beta}}  - {F}_{\alpha \delta}
{\Gamma}^{\delta}_{\beta \gamma} - {F}_{\delta \beta}
{\Gamma}^{\delta}_{\alpha \gamma};
\end{multline*}

Collect like terms in expression \verb|maxwell1|:
\begin{verbatim}
@canonicalise!(%);
\end{verbatim}
\begin{verbatim}
@collect_terms!(%);
\end{verbatim}
\begin{equation*}
maxwell1:= {\partial}_{\alpha}{{F}_{\beta \gamma}}  -
{\partial}_{\beta}{{F}_{\alpha \gamma}}  +
{\partial}_{\gamma}{{F}_{\alpha \beta}} ;
\end{equation*}
Thus, we obtain exactly equation~\eqref{eq:m:tensor:2:partial}.

Now, write Maxwell's equations~\eqref{eq:m:tensor:2}
and~\eqref{eq:m:tensor} in a medium in Cartesian coordinates with a
metric tensor $\eta_{\crd{\alpha} \crd{\beta}} =
\mathrm{diag}(1,-1,-1,-1)$:
\begin{equation}
  \label{eq:maxwell:media:decart}
  \begin{gathered}
    \partial_{{\alpha}} F_{{\beta}{\gamma}} +
    \partial_{{\beta}} F_{{\gamma}{\alpha}} +
    \partial_{{\gamma}} F_{{\alpha}{\beta}} = 0,
    \\
    \partial_{{\alpha}} H^{{\alpha}{\beta}} = \frac{4 \pi}{c}
    j^{{\beta}}.
  \end{gathered}
\end{equation}

Similarly, write Maxwell's equations in a vacuum in the effective
Riemannian space with a metric tensor $g_{\alpha \beta}$:
\begin{equation}
  \label{eq:maxwell:vacuum:riemann}
  \begin{gathered}
    \partial_{{\alpha}} f_{{\beta}{\gamma}} +
    \partial_{{\beta}} f_{{\gamma}{\alpha}} +
    \partial_{{\gamma}} f_{{\alpha}{\beta}} = 0,
    \\
    \frac{1}{\sqrt{-g}}
    \partial_{{\alpha}} \left( \sqrt{-g}
      h^{{\alpha}{\beta}} \right) = \frac{4 \pi}{c}
    \tilde{j}^{{\beta}}.
  \end{gathered}
\end{equation}
Here, $f_{\alpha \beta}$ and $h^{\alpha \beta}$ are the Maxwell and Minkowski tensors in the effective
Riemannian space.\footnote{Uppercase letters denote quantities in the
  Minkowski space, while lowercase letters denote quantities in the
  Riemannian space.}

For this purpose, in the Cadabra system, covariant divergence is
introduced:
\begin{verbatim}
div:=\nabla_{\alpha}A?^{\alpha \beta}-> 
  1/\sqrt{-g}\partial_{\alpha}
  (\sqrt{-g} A?^{\alpha \beta});
\end{verbatim}
\begin{equation*}
div:= {\nabla}_{\alpha}  {A?}^{\alpha \beta} \rightarrow \frac{1}{\sqrt{-g}} {\partial}_{\alpha}{(\sqrt{-g} {A?}^{\alpha \beta})};
\end{equation*}
Here, again, the postfix ``?'' converts the preceding symbol into a pattern.

Equation~\eqref{eq:m:tensor} is written in the effective Riemannian space:
\begin{verbatim}
riman:=\nabla_{\alpha} h^{\alpha \beta}=
  j^{\beta} 4 \pi / c;
\end{verbatim}
\begin{equation*}
riman:= {\nabla}_{\alpha}  {h}^{\alpha \beta} = 4 {j}^{\beta} \frac{\pi}{c};
\end{equation*}

Substituting the covariant divergence into equation \verb|riman| gives
equation~\eqref{eq:maxwell:vacuum:riemann}.
\begin{verbatim}
@substitute!(riman)(@(div));
\end{verbatim}
\begin{equation*}
riman:= \frac{1}{\sqrt{-g}} {\partial}_{\alpha}{(\sqrt{-g} {h}^{\alpha \beta})}  = 4 {j}^{\beta} \frac{\pi}{c};
\end{equation*}

Since the relation 
\begin{equation}
  \label{eq:f-g:relation}
  f_{\alpha \beta} = h_{\alpha \beta},
\end{equation}
holds for the vacuum, then, by raising the indices, we obtain
\begin{equation}
  \label{eq:f^-g:relation}
  f^{\alpha \beta} = g^{\alpha \gamma} g^{\beta \delta}
  h_{\gamma \delta}.
\end{equation}

For this purpose, in the Cadabra, the corresponding substitution is performed:
\begin{verbatim}
fh:=h^{\alpha \beta} -> 
  g^{\alpha \gamma}g^{\beta \delta} 
  f_{\gamma \delta};
\end{verbatim}

\begin{equation*}
fh:= {h}^{\alpha \beta} \rightarrow {g}^{\alpha \gamma} {g}^{\beta \delta} {f}_{\gamma \delta};
\end{equation*}

\begin{verbatim}
@substitute!(riman)(@(fh));
\end{verbatim}

\begin{equation*}
riman:= \frac{1}{\sqrt{-g}} {\partial}_{\alpha}{(\sqrt{-g} {g}^{\alpha \delta} {g}^{\beta \gamma} {f}_{\delta \gamma})}  = 4 {j}^{\beta} \frac{\pi}{c};
\end{equation*}

From term by term comparison, we have
\begin{gather}
  \label{eq:decart-riemann:relation}
  F_{\alpha \beta} = f_{\alpha \beta}, \qquad j^{\alpha} =
  \sqrt{-g} j^{\alpha},
  \\
  H^{\alpha \beta} = \sqrt{-g} g^{\alpha \gamma} g^{\beta \delta}
  F_{\gamma \delta}.
  \label{eq:decart-riemann:relation:g-f}
\end{gather}

Therefore, we obtain the relation
\begin{equation}
  \label{eq:f-durch-g}
  F_{\alpha \beta} = 
  \frac{1}{\sqrt{-g}} g_{\alpha \gamma} g_{\beta \delta} 
  H^{\gamma \delta}.
\end{equation}

Based on relation~\eqref{eq:f-durch-g}, an electric displacement
vector is expressed explicitly: 
\begin{equation}
  \label{eq:d^i}
  D^{i} = - \frac{\sqrt{-g}}{g_{0 0}} g^{i j} E_{j} + 
  \frac{1}{g_{0 0}} \varepsilon^{i j k} g_{j 0} H_{k}.
\end{equation}

In this case, the geometrized dielectric permittivity is written as
\begin{equation}
  \label{eq:e_ij} 
  \varepsilon^{i j} = - \frac{\sqrt{-g}}{g_{0 0}} g^{i j}.
\end{equation}

Similar manipulations are performed for a magnetic displacement vector.

Thus, we obtain the magnetic displacement vector
\begin{equation}
  \label{eq:b^i}
  B^{i} = - \frac{\sqrt{-g}}{g_{0 0}} g^{i j} H_{j} - 
  \frac{1}{g_{0 0}} \varepsilon^{i j k} g_{j 0} E_{k}
\end{equation}
and the geometrized magnetic permittivity
\begin{equation}
  \label{eq:mu_ij} 
  \mu^{i j} = - \frac{\sqrt{-g}}{g_{0 0}} g^{i j}.
\end{equation}

\section{Case study}
\label{sec:maxwell:calc}

The symbolic manipulations performed in the Cadabra system yield the
result that can be directly used for computer modeling and designing
of Maxwell (tensor-vector) optics elements, which can be represented
in curvilinear coordinates. To carry out such computations, we employ
the FORM system. 

As an example, we consider the popular application of transformation
optics: invisibility cap (or invisibility
cloak)~\cite{leonhardt:2009:light}.

\subsection{Invisibility cloak}

\begin{figure}[htb]
  \centering
  \includegraphics[width=0.8\linewidth]{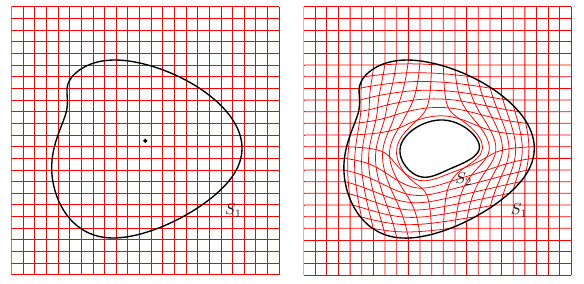}
  \caption{Two-dimensional projection of the coordinate transformation
    for the invisibility cloack}
  \label{fig:cloack}
\end{figure}

Here, we need to solve an inverse problem, i.e., find medium
parameters from the given coordinates related to the configuration of
the system. Let an object to be hidden be inside the region $S_1$  (that is
the invisibility cloak being designed) and be denoted by a point (see
figure~\ref{fig:cloack}). Originally, we have the Cartesian coordinate
system $x^{i'}$. Surround the object by the boundary $S_2$ and deform the coordinates in
the region between $S_1$ and $S_2$  by converting  into $x^{i'}$ в
$x^{i}$. Geodetic data outside $S_1$
remain unchanged. Moreover, light does not enter the region $S_2$ (any
object inside the region $S_2$ causes no changes outside the region
$S_1$). Hence, this structure actually acts as an invisibility cloack. 

For simplicity, we design a cylindrical invisibility cap. The original
flat\footnote{A manifold with zero curvature is called flat;
  therefore, the coordinate system superimposed on such a manifold is
  also called flat.}
coordinate system is denoted by $(r' , \varphi' , z')$. In these
coordinates, the
boundary $S_1$ is defined as $r' = b$. By transforming the coordinates $(r', \varphi' , z') \to (r,
\varphi, z)$, we obtain 
\begin{equation}
r' = 
\begin{dcases}
\frac{b−a}{a} r + a, & r' \leqslant b \\
r, & r' > b
\end{dcases}, 
\quad
\varphi' = \varphi, 
z' = z.
\end{equation}

Thus, the region $r' \leqslant b$ is contracted into the region $a \leqslant
r \leqslant b$, while the metric
tensor in the curvilinear coordinates takes the form $g_{ij} =
\mathrm{diag} [b^2/(b-a)^2, 1, 1]$. 

Below, we illustrate the calculation of the determinant for the metric
tensor $g$ in the FORM system. 

Disable the additional information about the computational process
(resources, time, etc.). To fit the calculation results in the paper,
decrease the output width up to 40 symbols (by default, it is 80):
\begin{verbatim}
Off statistic;
Format 40;
\end{verbatim}

Specify explicitly that the four-dimensional space is used. Then,
define basic elements, i.e., indices ($i$, $j$, $k$, and
$l$),\footnote{For compactness, we use Latin letters (instead of Greek
  letters) as FORM indices.} tensors ($g_{\alpha \beta}$) and
objects with no additional semantics ($a$, $b$):
\begin{verbatim}
Dimension 4;
Indices i, j, k, l;
Tensors g;
Symbols a,b;
\end{verbatim}

Define a function for the determinant (since there is no such a
function in the FORM). For this purpose, the Levi--Civita symbol is
used:
\begin{equation}
  \label{eq:FORM:detG}
  \det{g_{\alpha \beta}} =:
  g = \varepsilon^{0123} \varepsilon^{\alpha \beta \gamma \delta} 
   g_{0 \alpha} g_{1 \beta} g_{2 \gamma} g_{3 \delta}.  
\end{equation}

\begin{verbatim}
Local detG = e_(0,1,2,3) * e_(i,j,k,l) 
   * g(0,i) * g(1,j) * g(2,k) * g(3,l);
contract;
Print;
.sort
\end{verbatim}

The result is represented in the following form:
\begin{verbatim}
detG =
   g(0,0)*g(1,1)*g(2,2)*g(3,3) - g(0
   ,0)*g(1,1)*g(2,3)*g(3,2) - g(0,0)
   *g(1,2)*g(2,1)*g(3,3) + g(0,0)*g(
   1,2)*g(2,3)*g(3,1) + g(0,0)*g(1,3
   )*g(2,1)*g(3,2) - g(0,0)*g(1,3)*
   g(2,2)*g(3,1) - g(0,1)*g(1,0)*g(2
   ,2)*g(3,3) + g(0,1)*g(1,0)*g(2,3)
   *g(3,2) + g(0,1)*g(1,2)*g(2,0)*g(
   3,3) - g(0,1)*g(1,2)*g(2,3)*g(3,0
   ) - g(0,1)*g(1,3)*g(2,0)*g(3,2)
    + g(0,1)*g(1,3)*g(2,2)*g(3,0) + 
   g(0,2)*g(1,0)*g(2,1)*g(3,3) - g(0
   ,2)*g(1,0)*g(2,3)*g(3,1) - g(0,2)
   *g(1,1)*g(2,0)*g(3,3) + g(0,2)*g(
   1,1)*g(2,3)*g(3,0) + g(0,2)*g(1,3
   )*g(2,0)*g(3,1) - g(0,2)*g(1,3)*
   g(2,1)*g(3,0) - g(0,3)*g(1,0)*g(2
   ,1)*g(3,2) + g(0,3)*g(1,0)*g(2,2)
   *g(3,1) + g(0,3)*g(1,1)*g(2,0)*g(
   3,2) - g(0,3)*g(1,1)*g(2,2)*g(3,0
   ) - g(0,3)*g(1,2)*g(2,0)*g(3,1)
    + g(0,3)*g(1,2)*g(2,1)*g(3,0);
\end{verbatim}

In the FORM system, particular values of tensor components are given
by pattern substitutions (in this case, $g_{\alpha \beta} = \mathrm{diag} [1,
-b^2/(b-a)^2, -1, -1]$). The postfix modifier ``?''
converts the preceding symbol into a pattern:
\begin{verbatim}
id g(0,0) = 1;
id g(1,1) = - b^2/(b-a)^2;
id g(i?,i?) = - 1;
id g(i?,j?) = 0;
Print;
.sort
.end
\end{verbatim}

Thus, we obtain the value of the determinant:
\begin{equation}
  \label{eq:FORM:detG:final}
  \det{g_{\alpha \beta}} =: g = - \frac{b^2}{(b-a)^2}.
\end{equation}

\begin{verbatim}
detG =
    - 1/(b^2 - 2*a*b + a^2)*b^2;
\end{verbatim}

Now, based on~\eqref{eq:e_ij} and~\eqref{eq:mu_ij}, we find the parameters of the medium: 
\begin{equation}
\label{eq:cloac:e-mu:cilindric}
\begin{gathered}
\varepsilon_r = \mu_r = \frac{r − a}{r}, \\
\varepsilon_{\varphi} = \mu_{\varphi} = \frac{r}{r -a}, \\
\varepsilon_{z} = \mu_{z} =  \left( \frac{b}{b -a} \right)^2 \frac{r − a}{r}.
\end{gathered}
\end{equation}

The medium parameters can be similarly geometrized for the
invisibility cap that has other symmetry characteristics (or even has
no explicit symmetry). For example, for a spherical invisibility cap,
the medium parameters are as follows: 
\begin{equation}
\label{eq:cloac:e-mu:spheric}
\begin{gathered}
\varepsilon_r = \mu_r = \left( \frac{r − a}{r} \right)^2 \frac{b}{b -a}, \\
\varepsilon_{\vartheta} = \mu_{\vartheta} = \frac{b}{b -a}, \\
\varepsilon_{\varphi} = \mu_{\varphi} =  \frac{b}{b -a}.
\end{gathered}
\end{equation}

\section{Conclusions}
\label{sec:conclusion}

There are a great number of applied scientific problems that require
symbolic computations of two types: for prototype development of a new
software product and for serial numerical and symbolic computations on
the already debugged software product. Therefore, dividing computer
algebra problems into interactive and noninteractive (batch) seems
quite reasonable. 

In this paper, as a model problem, the geometrization of material
Maxwell's equations is considered. Our approach is based on the idea
of adopting the most promising mathematical and conceptual frameworks
from other scientific fields. In this case, the geometric paradigm is
used in the framework of the field theory. 

Since differential geometry forms the mathematical basis of the
geometric paradigm~\cite{wheeler:neutrinos::en}, we select the computer algebra systems that
are oriented to the tensor and vector analysis and support
manipulations with abstract tensors. Thus, the Cadabra system is used
for interactive operations, while the FORM system is employed for
batch computations. 

We hope that the selected examples clearly demonstrate the
possibilities of this approach for solving such problems.

\section{Acknowledgements}
\label{sec:ack}

We thank Jos Vermaseren for drawing the author's attention to the FORM
computer algebra system. 

This work was supported in part by the Russian Foundation for Basic
Research (project nos. 14-01-00628 and 15-07-08795). 

The computations were carried out on the Felix computational cluster
(Peoples' Friendship University of Russia) and on the HybriLIT
heterogeneous cluster (Multifunctional center for data storage,
processing, and analysis at the Joint Institute for Nuclear
Research). 

\bibliographystyle{elsarticle-num}

\bibliography{bib/geom-maxwell-cas/main,bib/geom-maxwell-cas/maxwell,bib/geom-maxwell-cas/form,bib/geom-maxwell-cas/self}

\end{document}


\graphicspath{{image/geom-maxwell-cas/}}

\title{Применение двух типов систем компьютерной алгебры для решения
  задач максвелловской оптики}

\author{Д. С. Кулябов}
\email{yamadharma@gmail.com}
\affiliation{Кафедра прикладной информатики и теории вероятностей,\\
  Российский университет дружбы народов,\\
  ул. Миклухо-Маклая, д.6, Москва, Россия, 117198}
\affiliation{Лаборатория информационных технологий,\\
Объединённый институт ядерных исследований,\\
ул. Жолио-Кюри 6, Дубна, Московская область, Россия, 141980}

\begin{abstract}
  
    






При решении задачи синтеза оптических систем на основе максвелловской оптики 
используется математический аппарат тензорного и векторного исчислений. Данный
математический аппарат подразумевает выполнение большого количества однотипных 
достаточно простых операций. Подобные операции очень хорошо поддерживаются 
системами компьютерной алгебры. Автор выделяет два этапа работы с математической
моделью: разработка модели и эксплуатация модели. Предполагается на каждом этапе 
использовать свою систему компьютерной алгебры. В качестве модельной задачи в 
статье рассматривается задача геометризации уравнений Максвелла. 
Для разных этапов
исследования были выбраны две системы компьютерной алгебры: 
Cadabra и FORM.

\end{abstract}

  \keywords{уравнения Максвелла; геометрическая парадигма;
    компьютерная алгебра; система компьютерной алгебры Cadabra;
    система компьютерной алгебры FORM}

\maketitle

\section{Введение}

В работе описано прикладное использование систем компьютерной алгебры
при решении задач проектирования максвелловской оптической системы,
описываемой уравнениями Максвелла в произвольных локально
ортогональных криволинейных координатах. Характер математического
описания задачи с использованием аппаратов тензорного и векторного
анализа формирует запрос на системы компьютерной алгебры со
встроенными функциями тензорного исчисления. Дополнительным
требованием является свободное распространение используемой системы
компьютерной алгебры в связи с тем, что на нынешнем этапе задача
представляет научный интерес и не является коммерчески
востребованной. Среди свободно распространяемых систем компьютерной
алгебры, имманентно реализующих аппарат тензорного и векторного
исчислений, для реализации двух этапов
исследования~\cite{hadamard:psychology::ru} выбирается две
разные системы.

Первый этап исследований заключается в написании прототипа
программы. Это успешно реализуется в системе Cadabra. После завершения
первого этапа исследований наступает второй этап, который имеет
исключительную ценность для специалиста-предметника, заинтересованного
в эксплуатации созданной программы. На этом этапе необходимо
проведение большого числа пробных расчётов с различными наугад или
планомерно изменяемыми параметрами с целью поиска наиболее интересного
из допустимых или даже нового решения прикладной предметной задачи. На
этом этапе максимально подходящей оказалась система FORM.

Попытка отыскать <<серебряную пулю>>~\cite{brooks:1986:silver_bullet}
не увенчались успехом, а поиски такой универсальной системы привели
автора к выводу, что каждая задача требует для своего решения своё
соответствующее средство.

Структура статьи следующая. В разделе~\ref{sec:notation} введены
основные обозначения и соглашения. В разделе~\ref{sec:cas-type}
рассматриваются типы систем компьютерной алгебры. В
разделе~\ref{sec:maxwell-geom} описывается формализм геометризации
уравнений Максвелла. Иллюстрации применения этого формализма на
примере задачи расчёта и компьютерного проектирования трёхмерных
волноводных объектов максвелловской оптики приведены
в разделе~\ref{sec:maxwell:calc}.

\section{Обозначения и соглашения}
\label{sec:notation}

  \begin{enumerate}

  \item В работе используется нотация абстрактных
    индексов~\cite{penrose-rindler:spinors::ru}. В данной нотации
    тензор как целостный объект обозначается просто индексом
    (например, $x^{i}$), компоненты обозначаются подчёркнутым индексом
    (например, $x^{\crd{i}}$).

  \item Будем придерживаться следующих соглашений.  Греческие индексы
    ($\alpha$, $\beta$) будут относиться к четырёхмерному пространству
    и в компонентном виде будут иметь следующие значения:
    $\crd{\alpha} = \overline{0,3}$. Латинские индексы из середины
    алфавита ($i$, $j$, $k$) будут относиться к трёхмерному
    пространству и в компонентном виде будут иметь следующие значения:
    $\crd{i} = \overline{1,3}$.

  \item Запятой в индексе обозначается частная производная по
    соответствующей координате ($f_{,i} := \partial_{i} f$); точкой с
    запятой --- ковариантная производная ($f_{;i} := \nabla_{i} f$).

  \item Для записи уравнений электродинамики в работе используется
    система СГС симметричная.
  \end{enumerate}

\section{Виды систем компьютерной алгебры}
\label{sec:cas-type}

Можно классифицировать системы компьютерной алгебры по разным
критериям. Нас же будет интересовать критерий
интерактивности. Первоначально на компьютерах преобладала пакетная
обработка. Однако, когда мощность компьютеров позволила сократить
время отклика до приемлемого значения, ведущей парадигмой стала
интерактивность. Каждая парадигма имеет свою область
применения. Интерактивность разработки была оставлена средствам
прототипирования, в то время как при создании рабочего продукта
использовались классический подход: написание текста программы,
компиляция, отладка и т.д. В системах символьных вычислений
интерактивность стала господствующим принципом. И действительно,
системы компьютерной алгебры были призваны повышать продуктивность
учёного, служили своего рода умным блокнотом. Задачи, где требовались
неинтерактивные вычисления были постепенно вытеснены на
обочину. Вместе с соответствующими системами компьютерной алгебры.

Архитектура многих современных систем компьютерной алгебры стала
ограничителем на пути повышения производительности
вычислений. Поэтому представляется актуальным подход, в рамках
которого используются разные системы компьютерной алгебры для разных
задач.

В работе предполагается рассмотреть два крайних полюса. С одной стороны~---
систему Cadabra, служащей для манипуляции с абстрактными
объектами. При этом подразумевается обязательная интерактивная работа
человека. С другой стороны~--- абсолютно неинтерактивная (даже
пакетная) система, похожая по циклу разработки скорее на классические
языки программирования, чем на обычные системы компьютерной алгебры.

\subsection{Cadabra}

Cadabra [\url{http://cadabra.phi-sci.com/}] относится к типу
специализированных систем компьютерной алгебры. Область её
специализации~--- теория поля. Поскольку сложные тензорные расчёты
являются неотъемлемой частью теории поля, неудивительно, что поддержка
тензорных расчётов находится в этой системе на высоком
уровне~\cite{kulyabov:2013:springer:cadabra,
  kulyabov:2009:nucleilett:cadabra,
  peeters:2006:field-theory_approach,
  peeters:2007:introducing_cadabra,
  peeters:2007:symbolic_field_theory, Brewin2010}.
В своей работе система Cadabra широко использует нотацию издательской
системы \TeX{}\footnote{Тесная интеграция Cadabra и \TeX{} привела к
  тому, что после установки дистрибутива системы \TeX{} \TeX{}Live-2015
  [\url{http://www.tug.org/texlive/}] Cadabra перестала работать с
  ошибкой \texttt{Undefined control sequence: \textbackslash{}int\_eval:w}. 
  Выяснилось,
  что причина ошибки в конфликте пакетов \texttt{breqn} и
  \texttt{expl3}, идущих в поставке \TeX{}Live-2015. Обновление пакета
  \texttt{breqn} из репозитория пакетов CTAN [\url{http://ctan.org/}]
  устранило проблему.}. 

На данный момент в системе Cadabra реализованы только операции с
абстрактными индексами.  Компонентные вычисления системой не
поддерживаются.  Впрочем, для реализации компонентных вычисления
потребуется дополнить систему Cadabra средствами компьютерной алгебры
общего назначения.

\subsection{Form}

Система компьютерной алгебры FORM [\url{http://www.nikhef.nl/~form/}]
очень сильно выделяется на фоне других систем компьютерной алгебры
направленностью не на интерактивную работу с пользователем, а на
пакетную обработку~\cite{tung:2005:FORM, vermaseren:2013:FORM_refman,
  heck:2000:formbook}. За счёт этого стало возможным снять такие
имманентные недостатки обычных систем компьютерной алгебры как высокая
требовательность к ресурсам, ограничения на объёмы вычислений,
медлительность. Система FORM поддерживает разные технологии
параллельных и распределённых вычислений~\cite{fliegner:1999:FORM},
такие как многонитевый подход, а также технологию MPI (несколько
реализаций). Система снабжена интерфейсом взаимодействия с внешними
программами~\cite{tentyukov:2007:extension_form}.  Всё это привело к
тому, что FORM зачастую используется в качестве бек-энда к другим (по
большей части интерактивным) системам компьютерной алгебры. Особенно
она популярна в квантово-полевых
вычислениях~\cite{boos:2010:automatic_calculation::ru,
  bunichev:2003:comphep+form, hahn:1999:FeynArts, hahn:2000:report,
  hahn:2007:FeynEdit}.

Система ведёт свою историю с 1984 года, однако открытой она стала лишь
в 2010~году.

Основные свойства FORM можно резюмировать следующим образом.
\begin{itemize}

\item Математические выражения произвольной длины (размеры
  ограничены только дисковым пространством).
  
\item Многопоточное исполнение, распараллеливание (MPI).
  
\item Быстрое вычисление следа ($\gamma$-матрицы).
  
\item Вывод в различных форматах (текст, Fortran).
  
\item Интерфейс для связи с внешними программами.

\end{itemize}

\subsection{Сравнение систем Cadabra и FORM}

Мы можем кратко суммировать необходимые свойства рассмотренных систем
компьютерной алгебры.

\paragraph{Cadabra}
\begin{itemize}
\item Основная особенность~--- естественная работа с
  тензорами. Изначально поддерживаются ковариантные и контравариантные
  индексы.
\item Эффективна для интерактивной работе для записи новых формул и
  соотношений.
\end{itemize}

\paragraph{FORM}
\begin{itemize}
\item Поддержка ковариантных и контравариантных индексов выполнена как
  искусственный приём.
\item Эффективна для финальных расчётов с уже известными формулами.
\end{itemize}

Таким образом, система Cadabra больше подходит для процесса написания
формул (рассмотрим это в разделе~\ref{sec:maxwell-geom}), для
непосредственных расчётов по уже выведенным формулам хорошо подойдёт
система FORM (см. раздел~\ref{sec:maxwell:calc}).

\section{Геометризация уравнений Максвелла}
\label{sec:maxwell-geom}

Две рассмотренные выше системы компьютерной алгебры применялись для
задачи компьютерных моделирования и синтеза элементов максвелловской
оптики в терминах тензорного исчисления и криволинейных координат.
И она естественно разбилась на две части. Вначале использовалась
система Cadabra, поскольку эта система и была разработана для задач
манипулирования тензорными объектами. 
Все манипуляции
проводились исключительно с абстрактными индексами. Использование
системы компьютерной алгебре на этом этапе позволило полностью
исключить бумагу и ручку из теоретической работы. На втором этапе
возникла, собственно, необходимость создания программного комплекса
для пакетного решения однотипных задач. Для этого была применена
система FORM. Данная система специализируется на
векторных и тензорных вычислениях.
Её спецификой является использование неинтерактивного (пакетного)
подхода, а также возможность использования распараллеливания
вычислений и внешней памяти.

\subsection{Идея геометризации}

Использованная методика моделирования и синтеза элементов
максвелловской оптики использует эффективную геометрическую
парадигму~\cite{wheeler:neutrinos::ru}. В
рамках этой парадигмы некоторые полевые параметры теории переводятся в
геометрические. В данном случае проводится геометризация
макроскопических параметров среды. Таким образом возможно решить
прямую и обратную задачи: по известным макроскопическим параметрам
определить траектории распространения электромагнитных волн; или же на
основании требуемых траекторий восстановить макроскопические параметры
среды.

Первые попытки применения методов дифференциальной геометрии в
электродинамике следует отнести к публикациям
И.~Е.~Тамма~\cite{tamm:1924:jrpc::ru, tamm:1925:jrpc::ru,
  tamm:1925:mathann}. В 1960~году Е.~Плебаньский предложил метод
геометризации материальных уравнений электромагнитного
поля~\cite{Plebanski1960, Felice1971, Leonhardt2008,
  leonhardt:2009:light}, ставший
классическим.

Основная идея наивной геометризации уравнений Максвелла заключается в следующем:
\begin{enumerate}
\item Записать уравнения Максвелла в среде в пространстве Минковского.
\item Записать вакуумные уравнения Максвелла в эффективном римановом
  пространстве.
\item Приравнять соответствующие члены уравнений.
\end{enumerate}
В результате мы получим выражение диэлектрической и магнитной
проницаемостей через метрику соответствующего эффективного
пространства.

Более подробно методика описана в статьях автора~\cite{kulyabov:2012:vestnik:2012-1,
  kulyabov:2013:springer:cadabra, kulyabov:2013:conf:maxwell,
  kulyabov:2011:vestnik:curve-maxwell::ru}.

\subsection{Общие соотношения}

Приведём основные сведения об уравнениях Максвелла.

Запишем уравнение Максвелла через тензоры
электромагнитного поля 
$F_{\alpha\beta}$ и
$H_{{\alpha}{\beta}}$~\cite{minkowski:1908,stratton:1948::ru}:
\begin{gather}
  \nabla_{{\alpha}} F_{{\beta}{\gamma}}+ \nabla_{{\beta}}
  F_{{\gamma}{\alpha}}+\nabla_{{\gamma}} F_{{\alpha}{\beta}} =
  F_{[\alpha \beta ; \gamma]} = 0,
  \label{eq:m:tensor:2}
  \\
  \nabla_{{\alpha}} H^{{\alpha}{\beta}}=\frac{4 \pi}{c}J^{{\beta}}.
  \label{eq:m:tensor}
\end{gather}
Здесь тензоры $F_{\alpha\beta}$ и $H^{\alpha\beta}$ имеют следующие
компоненты:
\begin{gather}
  F_{\crd{\alpha}\crd{\beta}}=
  \begin{pmatrix}
    0 & {E}_1 & {E}_2 & {E}_3 \\
    -{E}_1 & 0 & -{B}^3 & {B}^2 \\
    -{E}_2 & {B}^3 & 0 & -{B}^1 \\
    -{E}_3 & -{B}^2 & {B}^1 & 0
  \end{pmatrix},
  \label{eq:f_ab}
  \\
  H^{\crd{\alpha}\crd{\beta}}=
  \begin{pmatrix}
    0 & -{D}^1 & -{D}^2 & -{D}^3 \\
    {D}^1 & 0 & -{H}_3 & {H}_2 \\
    {D}^2 & {H}_3 & 0 & -{H}_1 \\
    {D}^3 & -{H}_2 & {H}_1 & 0
  \end{pmatrix}.
  \label{eq:g^ab}
\end{gather}

Для учёта среды введём макроскопические уравнения:
\begin{equation}
  \label{eq:constraint}
  D^{i} = \varepsilon^{i j} E_{j}, \qquad B^{i} = \mu^{i j} H_{j},
\end{equation}
\begin{equation}
  \label{eq:constraint:4}
  H^{\alpha \beta} = \lambda^{\alpha \beta}_{\gamma \delta} F^{\gamma
    \delta}, 
  \quad 
  \lambda^{\alpha \beta}_{\gamma \delta} = \lambda^{[\alpha \beta]}_{[\gamma \delta]}.
\end{equation}

Естественно, что в вакууме (при отсутствии среды)
соотношения~\eqref{eq:constraint} и~\eqref{eq:constraint:4}
приобретают следующий вид:
\begin{equation}
  \label{eq:perv:vac}
  \varepsilon^{i j} := \delta^{i j}, \qquad \mu^{i j} := \delta^{i j}.
\end{equation}
\begin{equation}
  \label{eq:constraint:vac}
  D^{i} = E^{i}, \qquad B^{i} = H^{i}, 
  \qquad H^{\alpha \beta} = F^{\alpha \beta}.
\end{equation}
  
\subsection{Геометризация уравнений Максвелла в декартовых
  координатах}

Заметим, что в
дифференциальном тождестве Бьянки~\eqref{eq:m:tensor:2} можно заменить
ковариантные производные на частные:
\begin{equation}
  \label{eq:m:tensor:2:partial}
  \begin{gathered}
    \nabla_{{\alpha}} F_{{\beta}{\gamma}}+ \nabla_{{\beta}}
    F_{{\gamma}{\alpha}}+\nabla_{{\gamma}} F_{{\alpha}{\beta}} =
    F_{[\alpha \beta ; \gamma]} = 0,
    \\
    \Downarrow
    \\
    \partial_{{\alpha}} F_{{\beta}{\gamma}}+ \partial_{{\beta}}
    F_{{\gamma}{\alpha}}+\partial_{{\gamma}} F_{{\alpha}{\beta}} =
    F_{[\alpha \beta , \gamma]} = 0.
  \end{gathered}
\end{equation}
Действительно:
\begin{multline}
  \label{eq:m:tensor:2:detail}
  \nabla_{{\alpha}} F_{{\beta}{\gamma}}+ \nabla_{{\beta}}
  F_{{\gamma}{\alpha}}+\nabla_{{\gamma}} F_{{\alpha}{\beta}} = {} \\
  {} =
  \partial_{{\alpha}} F_{{\beta}{\gamma}} -
  \Gamma_{\alpha\beta}^{\delta} F_{\delta \gamma} - \Gamma_{\alpha
    \gamma}^{\delta} F_{\beta \delta} + {} \\ {} +
  \partial_{{\beta}} F_{{\gamma}{\alpha}} - \Gamma_{\beta
    \gamma}^{\delta} F_{\delta \alpha} - \Gamma_{\beta
    \alpha}^{\delta} F_{\gamma \delta} + {} \\ {} +
  \partial_{{\gamma}} F_{{\alpha}{\beta}} - \Gamma_{\gamma
    \alpha}^{\delta} F_{\delta \beta} - \Gamma_{\gamma
    \beta}^{\delta} F_{\alpha \delta}.
\end{multline}

Продемонстрируем вычисление выражения~\eqref{eq:m:tensor:2:detail} в системе
компьютерной алгебры Cadabra\footnote{Точка в конце команды подавляет вывод
  результата, а точка с запятой заставляет напечатать результат.}. 

Зададим список индексов:
\begin{verbatim}
{\alpha,\beta,\gamma,\delta}::Indices(vector).
\end{verbatim}

Введём частную и ковариантную производные\footnote{Знак \# является
шаблоном для любого выражения.}:
\begin{verbatim}
\partial_{#}::PartialDerivative.
\end{verbatim}
\begin{verbatim}
\nabla_{#}::Derivative.
\end{verbatim}

При выводе выражения~\eqref{eq:m:tensor:2:detail} для нас будут иметь
значения симметрийные свойства символов Кристоффеля, которые зададим
с помощью диаграммы Юнга~\cite{fulton:young_tableaux::ru}: 
\begin{verbatim}
\Gamma^{\alpha}_{\beta \gamma}::
  TableauSymmetry(shape={2}, indices={1,2}).
\end{verbatim}

Для тензора $F_{\alpha \beta}$ нет необходимости приводить диаграмму
Юнга, достаточно указать, что он антисимметричный.
\begin{verbatim}
F_{\alpha \beta}::AntiSymmetric.
\end{verbatim}

Запишем выражение для ковариантной производной в виде
подстановки\footnote{Выражение $:=$ служит для задании метки.}:
\begin{verbatim}
nabla:=\nabla_{\gamma} A?_{\alpha \beta} -> 
  \partial_{\gamma}{A?_{\alpha \beta}} - 
  A?_{\alpha \delta}
  \Gamma^{\delta}_{\beta \gamma} - 
  A?_{\delta \beta}
  \Gamma^{\delta}_{\alpha \gamma};
\end{verbatim}
\begin{equation*}
  nabla:= {\nabla}_{\gamma}  {A?}_{\alpha \beta} \rightarrow
  ({\partial}_{\gamma}{{A?}_{\alpha \beta}}  - {A?}_{\alpha \delta}
  {\Gamma}^{\delta}_{\beta \gamma} - {A?}_{\delta \beta}
  {\Gamma}^{\delta}_{\alpha \gamma});
\end{equation*}
Мы используем постфикс <<?>>, превращающий
предыдущую букву в шаблон (в противном случае подстановка могла
быть использована лишь с фиксированной переменной). Заметим, что
применение модификатора <<?>> для индексов необязательно.

Теперь запишем уравнение~\eqref{eq:m:tensor:2}:
\begin{verbatim}
maxwell1:= \nabla_{\alpha}F_{\beta \gamma} + 
  \nabla_{\beta} F_{\gamma \alpha} + 
 \nabla_{\gamma} F_{\alpha \beta};
\end{verbatim}
\begin{equation*}
maxwell1:= {\nabla}_{\alpha}  {F}_{\beta \gamma} +
{\nabla}_{\beta}  {F}_{\gamma \alpha} + {\nabla}_{\gamma}
{F}_{\alpha \beta};
\end{equation*}

Выполним подстановку \verb|nabla| в выражение \verb|maxwell1|:
\begin{verbatim}
@substitute!(maxwell1)(@(nabla));
\end{verbatim}
\begin{multline*}
maxwell1:= {\partial}_{\alpha}{{F}_{\beta \gamma}}  - {F}_{\beta
  \delta} {\Gamma}^{\delta}_{\gamma \alpha} - {F}_{\delta \gamma}
{\Gamma}^{\delta}_{\beta \alpha} 
+ {} \\ {} + 
{\partial}_{\beta}{{F}_{\gamma
    \alpha}}  - {F}_{\gamma \delta} {\Gamma}^{\delta}_{\alpha \beta} -
{F}_{\delta \alpha} {\Gamma}^{\delta}_{\gamma \beta} 
+ {} \\ + {}
{\partial}_{\gamma}{{F}_{\alpha \beta}}  - {F}_{\alpha \delta}
{\Gamma}^{\delta}_{\beta \gamma} - {F}_{\delta \beta}
{\Gamma}^{\delta}_{\alpha \gamma};
\end{multline*}

В выражении \verb|maxwell1| приведём подобные:
\begin{verbatim}
@canonicalise!(%);
\end{verbatim}
\begin{verbatim}
@collect_terms!(%);
\end{verbatim}
\begin{equation*}
maxwell1:= {\partial}_{\alpha}{{F}_{\beta \gamma}}  -
{\partial}_{\beta}{{F}_{\alpha \gamma}}  +
{\partial}_{\gamma}{{F}_{\alpha \beta}} ;
\end{equation*}
В результате мы получили в точности
уравнение~\eqref{eq:m:tensor:2:partial}.

Теперь запишем уравнения Максвелла~\eqref{eq:m:tensor:2}
и~\eqref{eq:m:tensor} в среде в декартовых координатах с метрическим
тензором $\eta_{\crd{\alpha} \crd{\beta}} = \mathrm{diag}(1,-1,-1,-1)$:
\begin{equation}
  \label{eq:maxwell:media:decart}
  \begin{gathered}
    \partial_{{\alpha}} F_{{\beta}{\gamma}} +
    \partial_{{\beta}} F_{{\gamma}{\alpha}} +
    \partial_{{\gamma}} F_{{\alpha}{\beta}} = 0,
    \\
    \partial_{{\alpha}} H^{{\alpha}{\beta}} = \frac{4 \pi}{c}
    j^{{\beta}}.
  \end{gathered}
\end{equation}

Аналогично запишем вакуумные уравнения Максвелла в эффективном
римановом пространстве с метрическим тензором $g_{\alpha \beta}$:
\begin{equation}
  \label{eq:maxwell:vacuum:riemann}
  \begin{gathered}
    \partial_{{\alpha}} f_{{\beta}{\gamma}} +
    \partial_{{\beta}} f_{{\gamma}{\alpha}} +
    \partial_{{\gamma}} f_{{\alpha}{\beta}} = 0,
    \\
    \frac{1}{\sqrt{-g}}
    \partial_{{\alpha}} \left( \sqrt{-g}
      h^{{\alpha}{\beta}} \right) = \frac{4 \pi}{c}
    \tilde{j}^{{\beta}}.
  \end{gathered}
\end{equation}
Здесь $f_{\alpha \beta}$ и $h^{\alpha \beta}$ суть тензоры Максвелла и
Минковского в эффективном римановом пространстве\footnote{Заглавные буквы будем использовать для величин в
  пространстве Минковского, строчные~--- для величин в римановом пространстве.}.

В Cadabra для этого введём ковариантную дивергенцию:
\begin{verbatim}
div:=\nabla_{\alpha}A?^{\alpha \beta}-> 
  1/\sqrt{-g}\partial_{\alpha}
  (\sqrt{-g} A?^{\alpha \beta});
\end{verbatim}
\begin{equation*}
div:= {\nabla}_{\alpha}  {A?}^{\alpha \beta} \rightarrow \frac{1}{\sqrt{-g}} {\partial}_{\alpha}{(\sqrt{-g} {A?}^{\alpha \beta})};
\end{equation*}
Здесь опять постфикс <<?>> превращает предыдущий символ в шаблон.

Уравнение~\eqref{eq:m:tensor} запишем для случая эффективного
риманового пространства:
\begin{verbatim}
riman:=\nabla_{\alpha} h^{\alpha \beta}=
  j^{\beta} 4 \pi / c;
\end{verbatim}
\begin{equation*}
riman:= {\nabla}_{\alpha}  {h}^{\alpha \beta} = 4 {j}^{\beta} \frac{\pi}{c};
\end{equation*}

Подстановка ковариантной дивергенции в уравнение с меткой \verb|riman|
даст нам уравнение~\eqref{eq:maxwell:vacuum:riemann}.
\begin{verbatim}
@substitute!(riman)(@(div));
\end{verbatim}
\begin{equation*}
riman:= \frac{1}{\sqrt{-g}} {\partial}_{\alpha}{(\sqrt{-g} {h}^{\alpha \beta})}  = 4 {j}^{\beta} \frac{\pi}{c};
\end{equation*}

Поскольку для вакуума выполняется соотношение:    
\begin{equation}
  \label{eq:f-g:relation}
  f_{\alpha \beta} = h_{\alpha \beta},
\end{equation}
то, подняв индексы, получим:
\begin{equation}
  \label{eq:f^-g:relation}
  f^{\alpha \beta} = g^{\alpha \gamma} g^{\beta \delta}
  h_{\gamma \delta}.
\end{equation}

В Cadabra для этого сделаем соответствующую подстановку:
\begin{verbatim}
fh:=h^{\alpha \beta} -> 
  g^{\alpha \gamma}g^{\beta \delta} 
  f_{\gamma \delta};
\end{verbatim}

\begin{equation*}
fh:= {h}^{\alpha \beta} \rightarrow {g}^{\alpha \gamma} {g}^{\beta \delta} {f}_{\gamma \delta};
\end{equation*}

\begin{verbatim}
@substitute!(riman)(@(fh));
\end{verbatim}

\begin{equation*}
riman:= \frac{1}{\sqrt{-g}} {\partial}_{\alpha}{(\sqrt{-g} {g}^{\alpha \delta} {g}^{\beta \gamma} {f}_{\delta \gamma})}  = 4 {j}^{\beta} \frac{\pi}{c};
\end{equation*}

Сделав почленное сравнение, получим:
\begin{gather}
  \label{eq:decart-riemann:relation}
  F_{\alpha \beta} = f_{\alpha \beta}, \qquad j^{\alpha} =
  \sqrt{-g} j^{\alpha},
  \\
  H^{\alpha \beta} = \sqrt{-g} g^{\alpha \gamma} g^{\beta \delta}
  F_{\gamma \delta}.
  \label{eq:decart-riemann:relation:g-f}
\end{gather}

Отсюда получаем соотношение:
\begin{equation}
  \label{eq:f-durch-g}
  F_{\alpha \beta} = 
  \frac{1}{\sqrt{-g}} g_{\alpha \gamma} g_{\beta \delta} 
  H^{\gamma \delta}.
\end{equation}

Из соотношения~\eqref{eq:f-durch-g} получаем явную запись для вектора
электрической индукции:
\begin{equation}
  \label{eq:d^i}
  D^{i} = - \frac{\sqrt{-g}}{g_{0 0}} g^{i j} E_{j} + 
  \frac{1}{g_{0 0}} \varepsilon^{i j k} g_{j 0} H_{k}.
\end{equation}

При этом геометризованная диэлектрическая проницаемость будет иметь
вид:
\begin{equation}
  \label{eq:e_ij} 
  \varepsilon^{i j} = - \frac{\sqrt{-g}}{g_{0 0}} g^{i j}.
\end{equation}

Проведём аналогичные выкладки и для вектора магнитной индукции.

Соответственно получим вектор магнитной индукции
\begin{equation}
  \label{eq:b^i}
  B^{i} = - \frac{\sqrt{-g}}{g_{0 0}} g^{i j} H_{j} - 
  \frac{1}{g_{0 0}} \varepsilon^{i j k} g_{j 0} E_{k}
\end{equation}
и геометризованную магнитную проницаемость
\begin{equation}
  \label{eq:mu_ij} 
  \mu^{i j} = - \frac{\sqrt{-g}}{g_{0 0}} g^{i j}.
\end{equation}

\section{Пример расчёта}
\label{sec:maxwell:calc}

Проделанные с использованием системы Cadabra символьные манипуляции
позволили получить результат, который можно непосредственно применить
для компьютерных моделирования и проектирования элементов систем
максвелловской (тензорно-векторной) оптики, допускающих простое
представление в криволинейных координатах.
Для данных расчётов мы выбираем другую систему. Предлагается
использовать для этого систему FORM.

В качестве конкретного примера мы рассмотрим проектирование популярного
приложения трансформационной оптики~--- шапку-невидимку (или
плащ-невидимку)~\cite{leonhardt:2009:light}.

\subsection{Шапка-невидимка}

\begin{figure}[htb]
  \centering
  \includegraphics[width=0.8\linewidth]{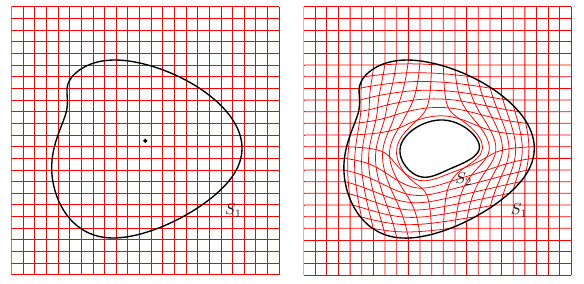}
  \caption{Двумерная проекция преобразования координат для
    шапки-невидимки}
  \label{fig:cloack}
\end{figure}

Итак, мы решаем обратную задачу, то есть по заданным координатам,
связанным с конфигурацией проектируемой системы, нам надо найти
параметры среды. Пусть объект, который надо скрыть, находится внутри
области $S_1$ (это и есть проектируемая шапка-невидимка) и обозначен
точкой на рисунке~\ref{fig:cloack}.  Первоначально будем иметь
декартову систему координат $x^{i'}$. Окружим наш объект границей
$S_2$ и деформируем координаты в области, ограниченной $S_1$ и $S_2$,
преобразовав $x^{i'}$ в $x^{i}$.  Геодезические за пределами $S_1$ не
изменились. Кроме того, свет проходит, не попадая в область $S_2$
(любой объект, помещённый внутри области $S_2$ не вызовет изменения за
пределами области $S_1$). То есть данная конструкция действительно
работает шапкой-невидимкой.

Для простоты будем проектировать пример шапки-невидимки цилиндрической формы.
Исходную плоскую\footnote{Плоским будем называть многообразие с
  нулевой кривизной. Соответственно, наложенную на это многообразие
  координатную систему будем называть плоской.} координатную систему
обозначим как $(r' , \varphi' , z')$. В
этих координатах границу $S_1$ зададим как $r' = b$.
Построим преобразование координат $(r', \varphi' , z') \to (r,
\varphi, z)$:
\begin{equation}
r' = 
\begin{dcases}
\frac{b−a}{a} r + a, & r' \leqslant b \\
r, & r' > b
\end{dcases}, 
\quad
\varphi' = \varphi, 
z' = z.
\end{equation}

В результате область $r' \leqslant b$ сжимается в область $a \leqslant
r \leqslant b$. Метрический тензор в криволинейных координатах
принимает вид $g_{ij} = \mathrm{diag} [b^2/(b-a)^2, 1, 1]$. 

Продемонстрируем вычисление определителя метрического тензора $g$ на
примере системы FORM.

Отключим дополнительную информацию по процессу вычисления (ресурсы,
время). Для того, чтобы вывод результатов уместился в статью, уменьшим
ширину вывода до 40 символов (по умолчанию 80).
\begin{verbatim}
Off statistic;
Format 40;
\end{verbatim}

Укажем явно, что мы работаем в 4-мерном пространстве. Далее зададим
основные элементы: индексы ($i$, $j$, $k$, $l$)\footnote{В качестве
  индексов в FORM будем использовать латинские буквы вместо греческих
  для большей компактности записи.}, тензоры ($g_{\alpha \beta}$) и
объекты, не имеющие
дополнительной семантики ($a$, $b$):
\begin{verbatim}
Dimension 4;
Indices i, j, k, l;
Tensors g;
Symbols a,b;
\end{verbatim}

Зададим формулу для определителя (поскольку в FORM такая функция не
задана). Для этого используем символ Леви-Чевиты:
\begin{equation}
  \label{eq:FORM:detG}
  \det{g_{\alpha \beta}} =:
  g = \varepsilon^{0123} \varepsilon^{\alpha \beta \gamma \delta} 
   g_{0 \alpha} g_{1 \beta} g_{2 \gamma} g_{3 \delta}.  
\end{equation}

\begin{verbatim}
Local detG = e_(0,1,2,3) * e_(i,j,k,l) 
   * g(0,i) * g(1,j) * g(2,k) * g(3,l);
contract;
Print;
.sort
\end{verbatim}

Результат будет представлен в следующем виде:
\begin{verbatim}
detG =
   g(0,0)*g(1,1)*g(2,2)*g(3,3) - g(0
   ,0)*g(1,1)*g(2,3)*g(3,2) - g(0,0)
   *g(1,2)*g(2,1)*g(3,3) + g(0,0)*g(
   1,2)*g(2,3)*g(3,1) + g(0,0)*g(1,3
   )*g(2,1)*g(3,2) - g(0,0)*g(1,3)*
   g(2,2)*g(3,1) - g(0,1)*g(1,0)*g(2
   ,2)*g(3,3) + g(0,1)*g(1,0)*g(2,3)
   *g(3,2) + g(0,1)*g(1,2)*g(2,0)*g(
   3,3) - g(0,1)*g(1,2)*g(2,3)*g(3,0
   ) - g(0,1)*g(1,3)*g(2,0)*g(3,2)
    + g(0,1)*g(1,3)*g(2,2)*g(3,0) + 
   g(0,2)*g(1,0)*g(2,1)*g(3,3) - g(0
   ,2)*g(1,0)*g(2,3)*g(3,1) - g(0,2)
   *g(1,1)*g(2,0)*g(3,3) + g(0,2)*g(
   1,1)*g(2,3)*g(3,0) + g(0,2)*g(1,3
   )*g(2,0)*g(3,1) - g(0,2)*g(1,3)*
   g(2,1)*g(3,0) - g(0,3)*g(1,0)*g(2
   ,1)*g(3,2) + g(0,3)*g(1,0)*g(2,2)
   *g(3,1) + g(0,3)*g(1,1)*g(2,0)*g(
   3,2) - g(0,3)*g(1,1)*g(2,2)*g(3,0
   ) - g(0,3)*g(1,2)*g(2,0)*g(3,1)
    + g(0,3)*g(1,2)*g(2,1)*g(3,0);
\end{verbatim}

Конкретные значения компонентов в FORM задаются с помощью подстановки
по шаблону (в данном случае $g_{\alpha \beta} = \mathrm{diag} [1,
-b^2/(b-a)^2, -1, -1]$). Постфиксный модификатор <<?>> превращает
предыдущий символ в шаблон:
\begin{verbatim}
id g(0,0) = 1;
id g(1,1) = - b^2/(b-a)^2;
id g(i?,i?) = - 1;
id g(i?,j?) = 0;
Print;
.sort
.end
\end{verbatim}

В результате, собственно, получаем значение определителя:
\begin{equation}
  \label{eq:FORM:detG:final}
  \det{g_{\alpha \beta}} =: g = - \frac{b^2}{(b-a)^2}.
\end{equation}

\begin{verbatim}
detG =
    - 1/(b^2 - 2*a*b + a^2)*b^2;
\end{verbatim}

Теперь, на основании~\eqref{eq:e_ij} и~\eqref{eq:mu_ij} получим параметры среды:
\begin{equation}
\label{eq:cloac:e-mu:cilindric}
\begin{gathered}
\varepsilon_r = \mu_r = \frac{r − a}{r}, \\
\varepsilon_{\varphi} = \mu_{\varphi} = \frac{r}{r -a}, \\
\varepsilon_{z} = \mu_{z} =  \left( \frac{b}{b -a} \right)^2 \frac{r − a}{r}.
\end{gathered}
\end{equation}

Аналогично, можно геометризованное представление параметров среды и в
случае других симметрий (или даже отсутствия какой-либо явной
симметрии) шапки-невидимки. Например, для
сферической шапки-невидимки параметры среды примут следующий вид:
\begin{equation}
\label{eq:cloac:e-mu:spheric}
\begin{gathered}
\varepsilon_r = \mu_r = \left( \frac{r − a}{r} \right)^2 \frac{b}{b -a}, \\
\varepsilon_{\vartheta} = \mu_{\vartheta} = \frac{b}{b -a}, \\
\varepsilon_{\varphi} = \mu_{\varphi} =  \frac{b}{b -a}.
\end{gathered}
\end{equation}

\section{Заключение}
\label{sec:conclusion}

Существует значительное число прикладных научных задач, для решения
которых необходимо реализовывать символьные вычисления двух типов:
прототипная разработка для нового программного продукта и проведение
серийных численных и символьных вычислений с использованием уже
отлаженного программного продукта. 
Поэтому разделение задач компьютерной алгебры на интерактивные и
неинтерактивные (пакетные) представляется автору вполне оправданным.

В качестве модельной задачи была выбрана задача геометризации
материальных уравнений Максвелла. В основе данного подхода лежит
идея об использовании наиболее перспективных
математического и понятийного аппарата из других областей науки. В
данном случае в рамках полевой теории используется геометрическая
парадигма.

Поскольку математической основой геометрической парадигмы является
дифференциальная геометрия~\cite{wheeler:neutrinos::ru}, в качестве
систем компьютерной алгебры 
выбирались системы, областью предметной ориентации которых являются
тензорное и векторное исчисления, с возможностью работы с абстрактными
тензорами. Соответственно, для интерактивной работы была выбрана
система Cadabra, а для пакетных расчётов~--- система FORM.

Автор надеется, что выбранные примеры демонстрируют возможность
применения данного подхода для таких задач.

\section{Благодарности}
\label{sec:ack}

Автор хотел бы выразить свою благодарность Йосу Вермасерену за то, что он обратил
моё внимание на систему компьютерной алгебры FORM.

Работа частично поддержана грантами РФФИ \textnumero 14-01-00628,
15-07-08795. 

Расчёты проведены на вычислительном кластере «Felix» РУДН и на
Гетерогенном вычислительном кластере «HybriLIT» Многофункционального
центра хранения, обработки и анализа данных ОИЯИ.

\bibliographystyle{gost2008}

\bibliography{bib/geom-maxwell-cas/main,bib/geom-maxwell-cas/maxwell,bib/geom-maxwell-cas/form,bib/geom-maxwell-cas/self}